# Title

***Output correction factors in a 1.5 T magnetic field determined at the central axis and at the lateral dose maximum for five detectors using alanine***

## Authors
Stephan Frick[1,†], Daniela Thorwarth[2,3], Thomas Hackel[1], Ralf-Peter Kapsch[1]

## Affiliations
[1] Physikalisch-Technische Bundesanstalt, Braunschweig, Germany
[2] Section for Biomedical Physics, Department of Radiation Oncology, University of Tübingen, Germany
[3] German Cancer Consortium (DKTK), partner site Tübingen, a partnership between DKFZ and University Hospital Tübingen, Germany

## [†]Corresponding author:

Stephan Frick
Physikalisch-Technische Bundesanstalt
Bundesallee 100, 38116 Braunschweig
E-Mail: Stephan.Frick@PTB.de/Stephan.Frick@gmx.de
Tel.: +49-(0)531-5926422

## Acknowledgement
This study received funding by the EMPIR program (Grant 19NRM01 MRgRT-DOS). We thank PTW Freiburg for providing microDiamond detectors for this work.

## Conflict of interest disclosure statement

**Word Count**

| | |
|---|---|
| Abstract: | 397 |
| Manuscript with captions/Tables: | 7087 |
| Manuscript without captions/Tables: | 6448 |
| Manuscript Figures/Tables number: | 7 |
| Supplementary Material pages: | 3 |

# Abstract

*Objective:*

During commissioning of an MR-linac, field output factors are measured utilizing output correction factors, $k_{\vec{B},Q_{\text{clin}},Q_{\text{msr}}}^{f_{\text{clin}},f_{\text{msr}}}$, which account for the influence of the field size and the magnetic field on the detector. In the literature to date, $k_{\vec{B},Q_{\text{clin}},Q_{\text{msr}}}^{f_{\text{clin}},f_{\text{msr}}}$ for 1.5 T have only been determined for the central axis (CAX), including our recent work.

At an MR-linac, the determination of the CAX position relies on e.g. MV-imaging and is at small field sizes located at the penumbra, whereas the maximum of the lateral profile (MAX) position is directly measurable. The aim of this study was to determine and compare $k_{\vec{B},Q_{\text{clin}},Q_{\text{msr}}}^{f_{\text{clin}},f_{\text{msr}}}$ at the CAX and MAX position.

*Approach:*

For this, $k_{\vec{B},Q_{\text{Ratio}}}$ - the influence of the magnetic field on the output correction factor without a magnetic field, $k_{Q_{\text{clin}},Q_{\text{msr}}}^{f_{\text{clin}},f_{\text{msr}}}$ - and $k_{\vec{B},Q_{\text{clin}},Q_{\text{msr}}}^{f_{\text{clin}},f_{\text{msr}}}$ were determined for both CAX and MAX fully experimentally for two MR-optimized ionization chambers, their conventional counterparts, and a solid-state detector.

Measurements were conducted using a water phantom in a mobile electromagnet positioned in front of a standard clinical linac (6 MV, 1.5 T). Profiles were measured with each detector to ensure accurate positioning at CAX/MAX. The uncertainty due to intra-type variation for tabulated $k_{\vec{B},Q_{\text{Ratio}}}$ and $k_{\vec{B},Q_{\text{clin}},Q_{\text{msr}}}^{f_{\text{clin}},f_{\text{msr}}}$ was assessed for the solid-state detector using four detectors. The change of absorbed dose to water was determined with alanine.

*Main results:*

Within the uncertainty, no difference in $k_{\vec{B},Q_{\text{Ratio}}}$ was observed between the MR-optimized chambers and their conventional counterparts. For the solid-state detector, no difference between the CAX and MAX position was found. For all detectors, $k_{\vec{B},Q_{\text{clin}},Q_{\text{msr}}}^{f_{\text{clin}},f_{\text{msr}}}$ remained constant down to a field size of 2.5 × 2.5 cm². At smaller field sizes, for the MAX position and for all detectors, $k_{\vec{B},Q_{\text{Ratio}}}$ decreased linearly. The reference point of the ionization chambers at the signal MAX position showed an average lateral displacement of 0.6 ± 0.1 mm over all field sizes compared to the dose MAX position, whereas the solid-state detector showed no displacement. The maximum standard uncertainty arising from intra-type variation was found to be 0.009 for the smallest field size, with and without a magnetic field.

*Significance:*

$k_{\vec{B},Q_{\text{Ratio}}}$ and $k_{\vec{B},Q_{\text{clin}},Q_{\text{msr}}}^{f_{\text{clin}},f_{\text{msr}}}$, including its uncertainty, were successfully determined for five detectors at CAX and MAX. For ionization chambers and small field sizes, $k_{\vec{B},Q_{\text{clin}},Q_{\text{msr}}}^{f_{\text{clin}},f_{\text{msr}}}$ is lower at MAX than at CAX, whereas for the solid-state detector, it remains the same. For the uncertainty of tabulated correction factors, intra-type variation must be considered.

# 1 Introduction

An MR-linac combines a magnetic resonance imaging (MRI) scanner with a linear accelerator (linac), enabling real-time imaging guidance with high tissue contrast (Liu et al., 2023). For commissioning the treatment planning system, the absorbed dose to water at small field sizes must be measured using suitable detectors (O'Brien et al., 2018). The small photon beam and the presence of a magnetic field in an MR-linac can influence the detector's signal, requiring correction with an output correction factor in magnetic fields, $k_{\vec{B},Q_{\text{clin}},Q_{\text{msr}}}^{f_{\text{clin}},f_{\text{msr}}}$.

In previous work (Frick et al., 2025), two methods for determining $k_{\vec{B},Q_{\text{clin}},Q_{\text{msr}}}^{f_{\text{clin}},f_{\text{msr}}}$ were presented and evaluated for detectors positioned at the central axis (CAX), with changes of absorbed dose to water being determined through Monte-Carlo simulation. Additionally, dose profiles were analysed, emphasizing the importance of defining a position for $k_{\vec{B},Q_{\text{clin}},Q_{\text{msr}}}^{f_{\text{clin}},f_{\text{msr}}}$ at small field sizes, since due to the shift and shape changes, CAX, the maximum (MAX) and the centre (midpoint between the 50% isodose positions) of the measured profile differ in position contrary to a situation without a magnetic field (see Figure 1).

In recent years, $k_{\vec{B},Q_{\text{clin}},Q_{\text{msr}}}^{f_{\text{clin}},f_{\text{msr}}}$ in a 1.5 T magnetic field for the off-maximum CAX position was investigated by several groups (Blum et al., 2021; Frick et al., 2025; Margaroni et al., 2025; Rojas-López et al., 2024; Yano et al., 2022). At an MR-linac, the MAX position is directly measurable with the detector, allowing for precise positioning. Contrary without turning the magnetic field off, the CAX position is not directly measurable with the detector but relies e.g. on MV-imaging. Data from both positions may be utilized for the MR-linac commissioning process. Since no data exist for MAX, it is essential to determine $k_{\vec{B},Q_{\text{clin}},Q_{\text{msr}}}^{f_{\text{clin}},f_{\text{msr}}}$ for MAX and comparing it to the CAX position, which allows to assess the effect of the different positions and evaluate whether one position offers advantages over the other.

Determining the absorbed dose to water using Monte Carlo simulations requires validation of the beam model, a complex process that may result in discrepancies between the simulated and the actual beam model. Experimental determination of the absorbed dose to water eliminates these discrepancies, although other uncertainties may arise depending on the detector used.

The use of alanine/EPR for determining the absorbed dose to water in a magnetic field has been studied in the literature for reference field sizes (Billas et al., 2020; Pojtinger et al., 2020). While (Billas et al., 2020) reported an influence of the magnetic field on alanine (1.0067 ± 0.0061 for 6 MV and 1.5 T), (Pojtinger et al., 2020) did not observe any effect, suggesting that the alanine holder might be responsible for this influence.

As shown and discussed in previous work (Frick et al., 2025), the factor $k_{\vec{B},Q_{\text{Ratio}}}$, which corrects for the influence of the magnetic field on the output correction factor without a magnetic field ($k_{Q_{\text{clin}},Q_{\text{msr}}}^{f_{\text{clin}},f_{\text{msr}}}$), is more consistent across the literature than $k_{\vec{B},Q_{\text{clin}},Q_{\text{msr}}}^{f_{\text{clin}},f_{\text{msr}}}$ itself. The latter discrepancy may be caused by a continuation of the big difference of $k_{Q_{\text{clin}},Q_{\text{msr}}}^{f_{\text{clin}},f_{\text{msr}}}$ in the literature (Hartmann and Zink, 2019) at small field sizes. $k_{\vec{B},Q_{\text{Ratio}}}$ only determines in the first step the change of perturbation factors and volume effect in a magnetic field relative to no magnetic field at a constant field size, which may be negligible for certain effects even in small fields, like the volume effect of the microDiamond detector (Cervantes et al., 2021).

Thus, this work aimed to determine $k_{\vec{B},Q_{\text{Ratio}}}$ and $k_{\vec{B},Q_{\text{clin}},Q_{\text{msr}}}^{f_{\text{clin}},f_{\text{msr}}}$ for two types of ionization chambers, their MR-optimized versions and a solid-state detector at both CAX and MAX for 6 MV and 1.5 T. The MR-optimized chambers lead to fewer artefacts in MR-imaging, which is important for the quality assurance of MR-based gating and tracking algorithms. The change in absorbed dose to water was determined fully experimentally with alanine/EPR. Furthermore, the uncertainty including intra-type variation was determined.

## 2  Material and Methods

### 2.1  Dosimetry formalism

For the determination of $k_{\vec{B},Q_{\text{clin}},Q_{\text{msr}}}^{f_{\text{clin}},f_{\text{msr}}}$, a method described in detail in our previous work (Frick et al., 2025) was used. In this method, a correction factor $k_{\vec{B},Q_{\text{Ratio}}}$ - accounting for the influence of the magnetic field on the output correction factor without a magnetic field ($k_{Q_{\text{clin}},Q_{\text{msr}}}^{f_{\text{clin}},f_{\text{msr}}}$) - is determined, as shown in the following equations:

$$k_{\vec{B},Q_{\text{clin}},Q_{\text{msr}}}^{f_{\text{clin}},f_{\text{msr}}} = k_{Q_{\text{clin}},Q_{\text{msr}}}^{f_{\text{clin}},f_{\text{msr}}} \cdot k_{\vec{B},Q_{\text{Ratio}}} \tag{1}$$

$$= k_{Q_{\text{clin}},Q_{\text{msr}}}^{f_{\text{clin}},f_{\text{msr}}} \cdot k_{\vec{B},M,Q_{\text{Ratio}}} \cdot c_{\vec{B}_{\text{Ratio}}} \tag{2}$$

$$= k_{Q_{\text{clin}},Q_{\text{msr}}}^{f_{\text{clin}},f_{\text{msr}}} \cdot \left[ \frac{M_{Q_{\text{clin}}}^{f_{\text{clin}}}}{M_{\vec{B},Q_{\text{clin}}}^{f_{\text{clin}}}} \middle/ \frac{M_{Q_{\text{msr}}}^{f_{\text{msr}}}}{M_{\vec{B},Q_{\text{msr}}}^{f_{\text{msr}}}} \right] \cdot \left[ \frac{D_{\vec{B},w,Q_{\text{clin}}}^{f_{\text{clin}}}}{D_{w,Q_{\text{clin}}}^{f_{\text{clin}}}} \middle/ \frac{D_{\vec{B},w,Q_{\text{msr}}}^{f_{\text{msr}}}}{D_{w,Q_{\text{msr}}}^{f_{\text{msr}}}} \right] \tag{3}$$

In this previous work, it was shown that $k_{\vec{B},Q_{\text{Ratio}}}$ is more comparable between literature than $k_{\vec{B},Q_{\text{clin}},Q_{\text{msr}}}^{f_{\text{clin}},f_{\text{msr}}}$ or $k_{Q_{\text{clin}},Q_{\text{msr}}}^{f_{\text{clin}},f_{\text{msr}}}$. To determine $k_{\vec{B},Q_{\text{clin}},Q_{\text{msr}}}^{f_{\text{clin}},f_{\text{msr}}}$, $k_{Q_{\text{clin}},Q_{\text{msr}}}^{f_{\text{clin}},f_{\text{msr}}}$, was taken from literature.

### 2.2  Experimental setup

Measurements were conducted at the Physikalisch-Technische Bundesanstalt (PTB, Germany) using an experimental setup which is identical to the one described in previous works (Frick et al., 2025, 2024). The setup consists of a water phantom placed within a mobile electromagnet, positioned in front of a clinical linear accelerator (Precise, Elekta AB, Sweden) with a nominal photon energy of 6 MV. The source-to-surface distance (SSD) was 110 cm and the detector's reference point was positioned at a water-equivalent depth of 10 cm.

The length of the equivalent square field $S_{\text{clin}}$ (Table 1) was determined using microDiamond (PTW, Germany) measurements and calculated following the protocols outlined in (BIR, 1996; TRS-483, 2017), as described in the previous work (Frick et al., 2025). Additionally, the reproducibility of $S_{\text{clin}}$ (Type-A uncertainty) was assessed through repeated measurements, incorporating data collected over the preceding year. All data was normalized to $S_{\text{clin}}$ = 4 cm by interpolating between the 3 × 3 and 4 × 4 cm² values.

Table 1: Length of the equivalent square field $S_{clin}$ for the nominal field sizes of the experimental setup

| Nominal field size [cm²] | 4 × 10 | 6 × 6 | 5 × 5 | 4 × 4 | 3 × 3 | 2 × 2 | 1.5 × 1.5 | 1 × 1 | 0.5 × 0.5 |
|---|---|---|---|---|---|---|---|---|---|
| $S_{\text{clin}}$ [cm] | 6.98 | 7.25 | 6.05 | 4.84 | 3.64 | 2.43 | 1.85 | 1.26 | 0.72 |

## 2.3 Detector measurements: $k_{\vec{B},M,Q_{\text{Ratio}}}$

$k_{\vec{B},M,Q_{\text{Ratio}}}$ was determined for a magnetic flux density of ±1.5 T for five detectors manufactured by PTW Freiburg (Germany): two MR-optimized ionization chambers - Semiflex 3D MR PTW31024 (sensitive volume 0.07 cm$^3$), PinPoint 3D MR PTW31025 (sensitive volume 0.016 cm$^3$) -, their conventional counterpart - Semiflex 3D PTW31021, PinPoint 3D PTW31022 (same geometry and volume) - and a solid-state detector (microDiamond PTW60019).

Axial and radial detector orientations were used. For the first, the detector's axis was aligned parallel to the beam axis and perpendicular to the magnetic field lines. This orientation was used for the microDiamond, PinPoint 3D, and PinPoint 3D MR. For the radial orientation, the detector's axis was aligned perpendicular to both the magnetic field lines and the beam axis. In this orientation, a negative $B$ value indicates that the initial Lorentz force $F_L$ acting on electrons travelling along the beam axis is directed toward the chamber tip. This orientation was used for the Semiflex 3D and Semiflex 3D MR.

For each detector and field size, lateral dose profiles were measured both without and with a 1.5 T magnetic field for nominal field sizes of ≤3 × 3 cm$^2$. In this work, the position of the profile's maximum without a magnetic field is referred to as CAX and the position of the profile's maximum with a magnetic field will be referred to as MAX. The profile's maximum was determined using a polynomial fit around the measurement point. No effective point of measurement (EPOM) shift was applied to the ionization chambers. Consequently, all data are referenced directly to the detectors reference point plane and remain independent of any EPOM displacement. Therefore, the MAX positions obtained with ionization chambers might be different due to a lateral shift of the EPOM observed in magnetic fields (Episkopakis et al., 2024; O'Brien et al., 2018). To investigate this, two shifts were analysed:

i. $EPOM_{L,Centre}$ shift: The difference between the detector and absorbed dose to water values of the lateral 0 T-centre-to-1.5 T-centre distance $d_{L,Centre}$. $d_{L,Centre}$ is the distance between the centre - defined as the midpoint between the 50% isodose positions of the profile - without (ideally at CAX) and with a 1.5 T magnetic field (cf. Figure 1). $d_{L,Centre}$ of the absorbed dose was determined with the microDiamond, since literature indicates it is representative of the dose shift (Episkopakis et al., 2024).

ii. $EPOM_{L,MAX}$ shift: The difference between the detector and absorbed dose to water values of the lateral 0 T-MAX-to-1.5 T-MAX distance $d_{L,MAX}$. $d_{L,MAX}$ is the distance between the maximum of the profile without (ideally at CAX) and with a 1.5 T magnetic field (cf. Figure 1). $d_{L,MAX}$ of the absorbed dose was determined with the microDiamond and checked with alanine values.

$k_{\vec{B},M,Q_{\text{Ratio}}}$ was determined at CAX and MAX for all detectors for nominal field sizes ranging from 6 × 6 cm$^2$ to 0.5 × 0.5 cm$^2$. Furthermore, for the ionization chambers, $k_{\vec{B},M,Q_{\text{Ratio}}}$ was measured at the MAX position determined with the microDiamond, which coincides with the absorbed dose to water maximum position (Episkopakis et al., 2024). An external monitor chamber, as described in (Krauss and Kapsch, 2014), whose signal is dependent of field size, was used to account for fluctuations in the linac's output. Additionally, the correction factor $k_{\text{pol}}$ for the polarity effect was measured according to (TRS-483, 2017) and applied with and without a magnetic field for every ionisation chamber type.

To determine reproducibility, measurements were repeated at least three times per detector type (MR-optimized and conventional models were assumed to have the same reproducibility).

For the final detector uncertainty estimation of $k_{\vec{B},M,Q_{\text{Ratio}}}$, calculated according to (GUM, 2008), the following were considered:

- the reproducibility (type-A uncertainty)
- a type-B uncertainty of 0.16% - estimated considering uncertainties for temperature and atmospheric pressure measurements (calculated based on information from the instrument manufacturer) as well as additional uncertainties for the electrometer –
- a type-B uncertainty (rectangle distribution) for the effect of the magnetic field on the setup (monitor chambers, asymmetrical field shape and hall probe)
- a Type-B uncertainty for a tabulated factor arising from intra-type variation

As described in the previous work (Frick et al., 2025), the type-B uncertainty arising from the influence of the magnetic field on the setup was estimated by considering the ±1.5 T results of axial-oriented detectors.

For the uncertainty arising from intra-type variation, four different detectors for the microDiamond and two Semiflex 3D chambers for the ionization chambers were used.

## 2.4 Absorbed dose to water measurement: $c_{\vec{B}_{\text{Ratio}}}$

The change in absorbed dose to water at CAX and MAX with and without a 1.5 T magnetic field was determined using alanine dosimeters. These measurements were used to calculate $c_{\vec{B}_{\text{Ratio}}}$ for equation (2).

Alanine dosimetry is based on the formation of stable free radicals by ionizing radiation, which is subsequently detected and quantified using electron spin resonance (EPR) spectroscopy. The alanine pellets consist of 90% L-α-alanine and 10% high melting point paraffin wax (Billas et al., 2020).

The internal monitor chamber (monitor units) was calibrated with and without a magnetic field by a Farmer type chamber (PTW30013) at reference field size. For this calibration, $k_{\vec{B},Q}$ values from (Pojtinger, 2021) were applied to correct the influence of the magnetic field on $k_Q$. To determine $c_{\vec{B}_{\text{Ratio}}}$ at CAX and MAX, for each field size lateral dose profiles were measured without and with the magnetic field using a microDiamond detector to calculate the shift between the room laser crosshair and CAX and MAX of the setup. Without moving the multileaf collimator or jaws of the linac, an in-house designed holder - housing a smaller, removable insert containing five cylindric alanine pellets (radius = 2.4 mm, height = 2.8 mm) - was positioned according to the room laser analogue to the microDiamond, with the central alanine pellet positioned on the laser crosshair. Afterwards, the calculated shifts for CAX or MAX were applied. Three measurements for each field size (with non-irradiated alanine pellets) were performed: one on CAX without a magnetic field, one on CAX with a magnetic field, and one on MAX with a magnetic field. Before irradiation, the holder was submerged in water for at least 15 minutes to allow the alanine pellets to reach thermal equilibrium with the water. This procedure was repeated for each field size.

For the alanine evaluation purposes, the central alanine pellet was irradiated with an approximate dose of 15 Gy for each field size, achieved by scaling the monitor units according to the detector output ratio measurements with the microDiamond. To evaluate the field output factors, the absorbed dose determined with the alanine was divided by the corresponding monitor units. Since these measurements spanned several days, the dose in the 4 × 4 cm² field without a magnetic field was measured daily to account for fluctuations in the setup between the days.

The procedure for evaluating the applied dose on the alanine pellets was done as described in detail by (Anton, 2006, 2005). The signal was corrected for photon energy, water temperature, the spatial resolution of the evaluation device and the volume averaging effect. For each pellet, the volume averaging effect was corrected using the factor $k_{\text{vol}}$, derived from profiles measured with the microDiamond. For measurements at MAX, a fitting function was applied to the alanine profiles to determine the dose maximum precisely.

For the uncertainty of $c_{\vec{B}_{\text{Ratio}}}$, a Type-A standard uncertainty was estimated by reproducibility per field size and variations of the five alanine pellets determined using the large 4 × 10 cm² field size. For the Type-B standard uncertainty, a value of 0.49% was used according to (Anton, 2006), accounting for effects such as the correction for temperature, radiation quality and variations of probe mass.

## 2.5 $k_{\vec{B},Q_{\text{Ratio}}}$ and $k_{\vec{B},Q_{\text{clin}},Q_{\text{msr}}}^{f_{\text{clin}},f_{\text{msr}}}$

$k_{\vec{B},Q_{\text{Ratio}}}$ was determined as the product of $k_{\vec{B},M,Q_{\text{Ratio}}}$ and $c_{\vec{B}_{\text{Ratio}}}$. For the calculation of $k_{\vec{B},Q_{\text{clin}},Q_{\text{msr}}}^{f_{\text{clin}},f_{\text{msr}}}$ according to equation (1), $k_{Q_{\text{clin}},Q_{\text{msr}}}^{f_{\text{clin}},f_{\text{msr}}}$ values including its uncertainty from the literature were utilized (Casar et al., 2020, 2019; TRS-483, 2017). The exact $k_{Q_{\text{clin}},Q_{\text{msr}}}^{f_{\text{clin}},f_{\text{msr}}}$ values for the field sizes used in this work were calculated using a fit described in (TRS-483, 2017). Since no data for the PinPoint 3D 31022 is available in (TRS-483, 2017), data of the previous model, PinPoint 3D 31016, was used. Additionally, as no $k_{Q_{\text{clin}},Q_{\text{msr}}}^{f_{\text{clin}},f_{\text{msr}}}$ values for the MR-optimized chambers exist, values from their non-MR-optimized counterparts (same chamber geometry) were applied.

In this work, the calculated correction factors $k_{\vec{B},Q_{\text{Ratio}}}$ and $k_{\vec{B},Q_{\text{clin}},Q_{\text{msr}}}^{f_{\text{clin}},f_{\text{msr}}}$ can be used to determine the absorbed dose to water and therefore the field output factor at two positions, which are at the lateral reference point plane of the detector at a fixed depth of 10 cm:

i. <u>CAX position</u>: At the central axis of the photon beam. Without a magnetic field, ideally this corresponds to the position of the maximum of the lateral dose profile. With a magnetic field and in small field sizes, the CAX position differs from the position of dose maximum and might even be located in the penumbra region (see Figure 1). At an MR-linac without turning the magnetic field off, the CAX position cannot be measured directly with a detector and must be determined indirectly (e.g. via MV-imaging). The reference point of the detector is at the same position in the water phantom as the position of the desired field output factor.

ii. <u>MAX position:</u> Position of the maximum of the lateral absorbed dose profile MAX$_{\text{dose}}$. At an MR-linac, the position of the maximum of the detector's signal MAX$_{\text{signal}}$ can be measured directly, allowing for precise positioning. For detectors with a lateral EPOM shift in magnetic fields (Episkopakis et al., 2024), the MAX$_{\text{dose}}$ position differs from the MAX$_{\text{signal}}$ position of the detector if the lateral EPOM shift is not considered (see Figure 1). In this case, the reference point of the detector at MAX$_{\text{signal}}$ is not at the same position in the water phantom as the position of the desired field output factor at MAX$_{\text{dose}}$.

Field output factor data from both positions may be used for the MR-linac commissioning process. Additionally for the ionization chambers, $k_{\vec{B},Q_{\text{Ratio}}}$ was determined with the detector's reference point at the MAX$_{\text{dose}}$ position (identified using the microDiamond). In this work, unless explicitly stated otherwise, correction factors for the MAX position refers to the detector's reference point placed at the MAX$_{\text{signal}}$ position. Note that both detector MAX positions determine the field output factor (absorbed dose to water) at MAX$_{\text{dose}}$.

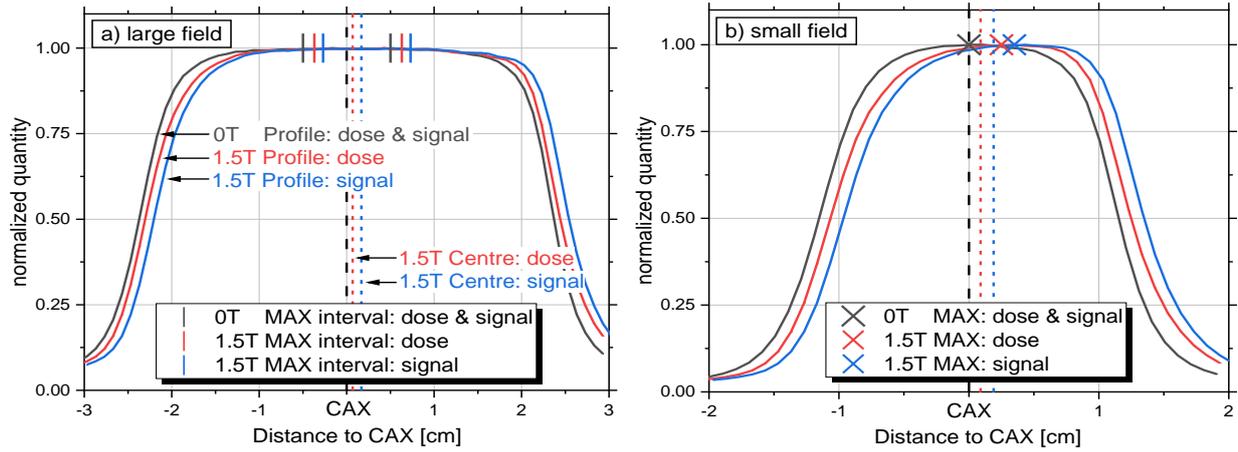

*Figure 1: Schematic lateral profiles of dose and detector signal (illustrating a lateral EPOM shift in a magnetic field) without (black) and with (red, blue) a magnetic field. Data adapted from previous work (Frick et al., 2025).*
*(a) For large fields like 4 × 4 cm², both the central axis (CAX) and the centre of all profiles remain coincident at the profiles maximum (MAX) interval.*
*(b) For smaller fields like 2 × 2 cm², in a magnetic field the profiles centre and CAX differ from MAX. For detectors with a lateral EPOM shift, the signal and dose MAX and centre are not at the same position.*

# 3 Results

## 3.1 General results

The Type-A standard uncertainty of the determined length of the equivalent square field $S_{\text{clin}}$, was found to have a maximum of 0.02 cm for the smallest field size, highlighting the excellent reproducibility of $S_{\text{clin}}$.

For the polarization correction factor $k_{\text{pol}}$, no difference between a magnetic field and no magnetic field was observed. At MAX, $k_{\text{pol}}$ remained constant over all field sizes for the Semiflex 3D types. For the PinPoint 3D types, it remained constant except for a 0.2% difference at the smallest field size.

The volume averaging effect for each alanine pellet was determined and corrected using $k_{\text{vol}}$ for all field sizes and measurement positions, both with and without a magnetic field. For the central pellet, $k_{\text{vol}}$ was 1.001 at 2 × 2 cm² and increased exponentially to 1.044 for 0 T and 1.045 for 1.5 T at both CAX and MAX at 0.5 × 0.5 cm², showing no big differences between $k_{\text{Vol}}$ with and without a magnetic field.

Within the uncertainty and in a magnetic field, no differences were observed in the maximum values of the alanine profiles between the CAX and MAX position.

In general, the shape and shift of the profiles measured with the microDiamond were consistent with those obtained from the alanine profile, as seen in Figure 2 and in Supplementary Figure S.1. Figure 2 further confirms this agreement, where the microDiamond lateral maximum shift $d_{L,MAX}$ (used for the MAX measurement) agree with the alanine data, meaning no need for a $EPOM_{L,MAX}$ shift. Compared to the microDiamond, the profiles of the ionization chambers in a magnetic field exhibit a slightly different shape of the penumbras and a larger $d_{L,MAX}$, resulting in a $EPOM_{L,MAX}$ shift of about 0.6 ± 0.1 mm, which was approximately constant over the field sizes. The lateral centre shift $d_{L,Centre}$ was also about constant over all field sizes, with a slight linear decrease for the microDiamond and PinPoint 3D types towards small field sizes. The $d_{L,Centre}$ difference between microDiamond and PinPoint 3D types ($EPOM_{L,Centre}$) mirrors the $EPOM_{L,MAX}$ (0.5 ± 0.1 mm), whereas for the Semiflex 3D types, $EPOM_{L,Centre}$ differed and was 1.1 ± 0.1 mm in stem direction and 0.9 ± 0.1 mm in tip direction.

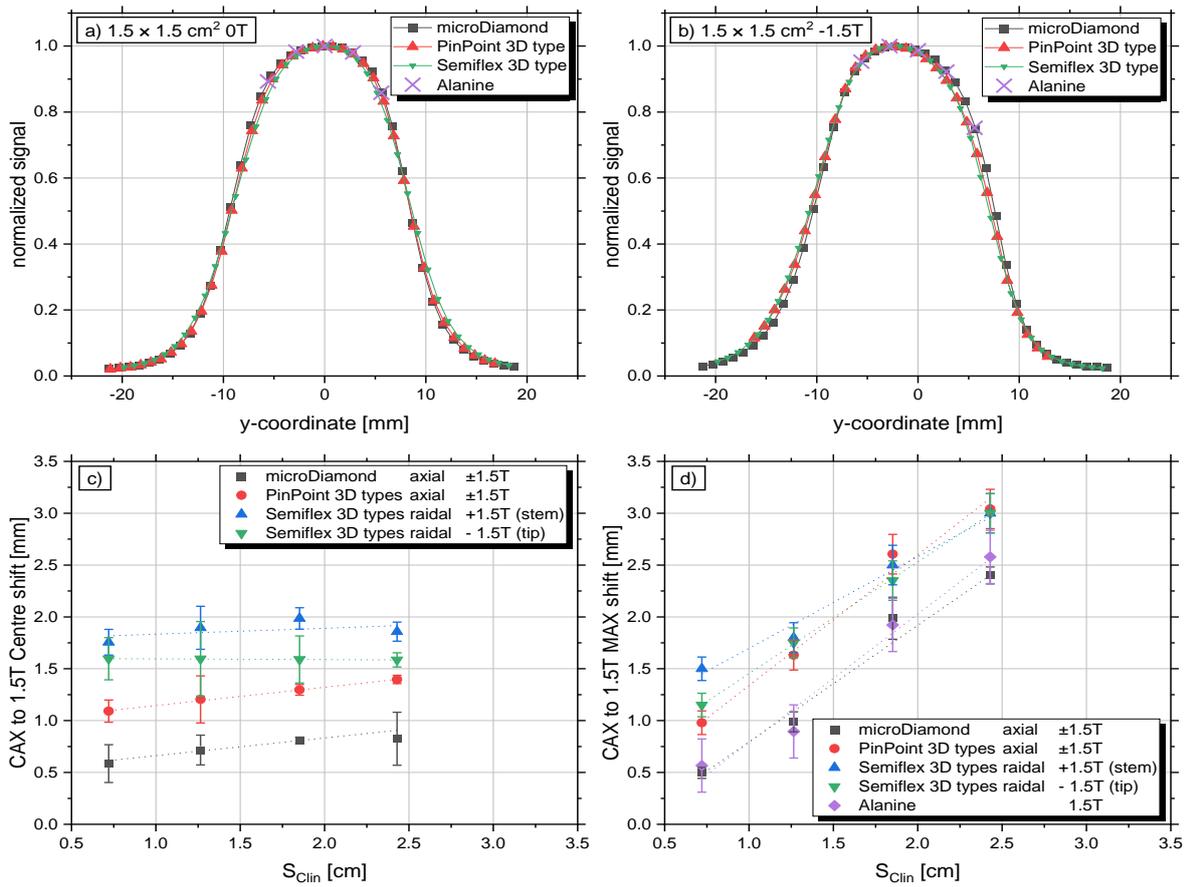

*Figure 2: Top row: Example of relative lateral dose profiles measured with different detectors in a) 0 T and b) –1.5 T. The zero y-coordinate represents the maximum at 0 T. c) $d_{L,Centre}$. The difference of $d_{L,Centre}$ between the ionization chamber and microDiamond is $EPOM_{L,Centre}$. d) $d_{L,MAX}$, which was used for the MAX position measurement. The difference of $d_{L,MAX}$ between the ionization chamber and microDiamond/alanine is $EPOM_{L,MAX}$. The data of the PinPoint 3D types includes PTW31022 and PTW31025, the data of the Semiflex 3D types includes PTW31021 and PTW31024. The intervals represent the standard uncertainty (rectangle distribution) and were calculated considering reproducibility, effect of the magnetic field on the setup (by comparing ±1.5 T data of the microDiamond and the PinPoint 3D types), and intra-type variation (for microDiamond, PinPoint 3D types, Semiflex 3D types).*

## 3.2 Detector data: $k_{\vec{B},M,Q_{\text{Ratio}}}$

In Figure 3, results for $k_{\vec{B},M,Q_{\text{Ratio}}}$ are shown. The data of the MR-optimized chambers and their conventional counterparts agree within the uncertainty. For all detectors, $k_{\vec{B},M,Q_{\text{Ratio}}}$ remains constant down to a field size of 3.5 × 3.5 cm² before increasing, except for the PinPoint 3D types at MAX, where $k_{\vec{B},M,Q_{\text{Ratio}}}$ remains approximately constant across all field sizes. For the ionization chambers positioned at the MAX$_{\text{dose}}$ position (measured by microDiamond), $k_{\vec{B},M,Q_{\text{Ratio}}}$ increased stronger.

The mean Type-A standard uncertainty (reproducibility) of $k_{\vec{B},M,Q_{\text{Ratio}}}$ across all field sizes was 0.004 for the microDiamond and 0.002 for the ionization chambers. This uncertainty increased for the ionization chambers at smaller field sizes to 0.007, while it remained constant for the microDiamond across all field sizes. No differences were observed between CAX and MAX.

The mean Type-B standard uncertainty for a tabulated factor due to intra-type variation across all field sizes was 0.007 for the microDiamond (n=4), with a maximum uncertainty of 0.009 at the smallest field size, showing no differences between CAX and MAX.
For the Semiflex 3D (n=2), the mean standard uncertainty was 0.004 across all field sizes and 0.007 at the smallest field size. At MAX, the uncertainty remained constant across all field sizes, whereas at CAX, it increased to 0.010 for the smallest field size.
The intra-type variation of detector output ratios was 0.002 (mean) and 0.009 (maximum) for the microDiamond and 0.003 (mean) and 0.005 (maximum) for the Semiflex 3D.

The mean Type-B standard uncertainty for the magnetic field's influence on the experimental setup across all field sizes was 0.005, with a maximum influence of 0.011 at the smallest field size.

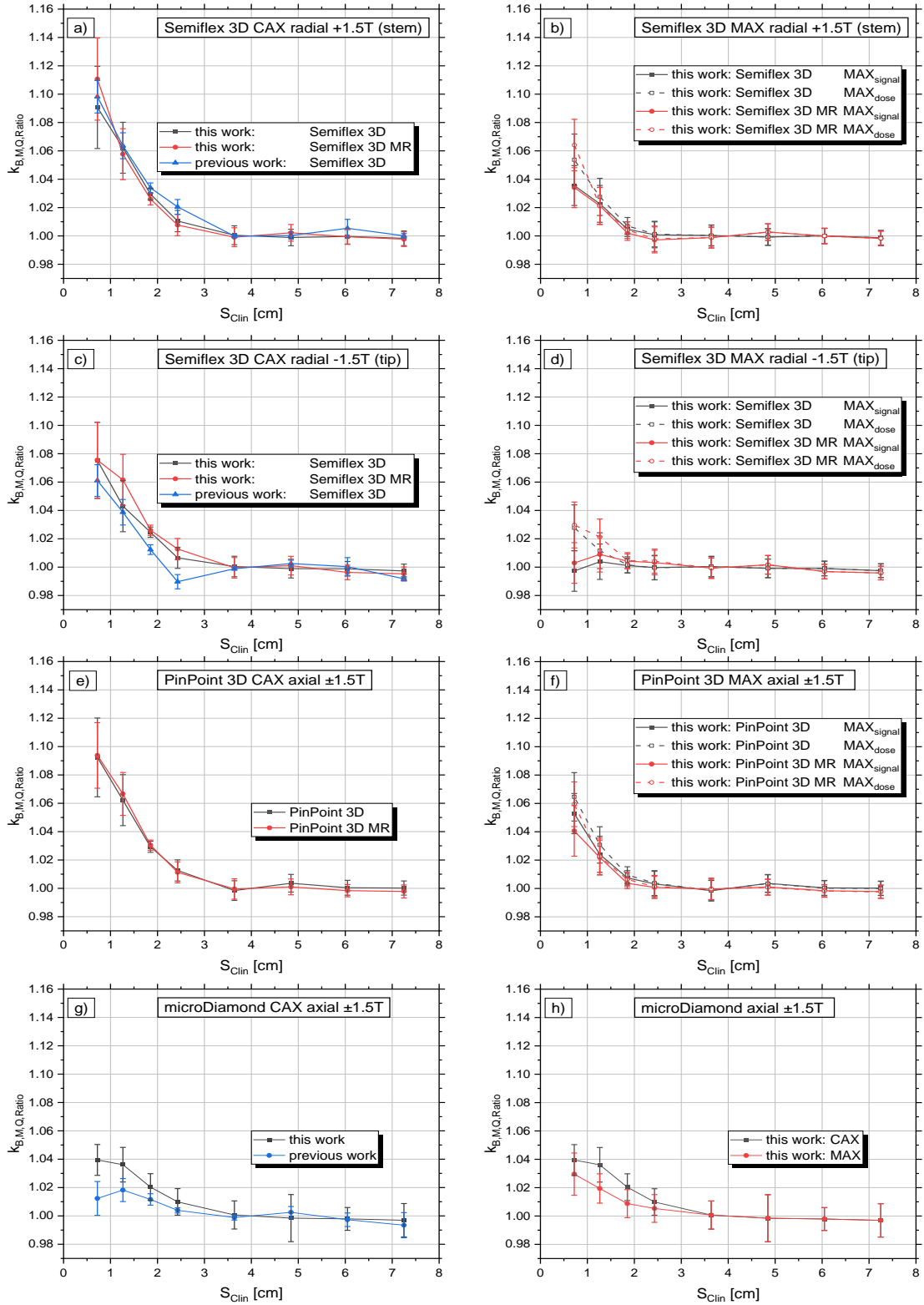

Figure 3: $k_{\vec{B},M,Q_{Ratio}}$ for different detectors and orientations for the central axis position (CAX, left column) and the profile maximum position (MAX, right column). For the ionization chambers at MAX, the reference point of the chamber was positioned at the profiles signal maximum $MAX_{signal}$ and at the absorbed dose maximum $MAX_{dose}$ (determined by the microDiamond). For the microDiamond, these positions are the same. The error bars indicate the standard uncertainty. Data (including uncertainty) marked as "previous work" is taken from (Frick et al., 2025).

## 3.3 Change of absorbed dose to water: $c_{\vec{B}_{\text{Ratio}}}$

In Figure 4, $c_{\vec{B}_{\text{Ratio}}}$ is presented, including simulated CAX data from the previous work (Frick et al., 2025), which agrees within the uncertainty with the measured data. For alanine, the total mean standard uncertainty across all field sizes for CAX and MAX was 0.006, increasing to 0.008 at the smallest field size. For both CAX and MAX, $c_{\vec{B}_{\text{Ratio}}}$ remains constant down to a field size of 2.5 × 2.5 cm², after which it decreases linearly, with a sharper drop observed at CAX.

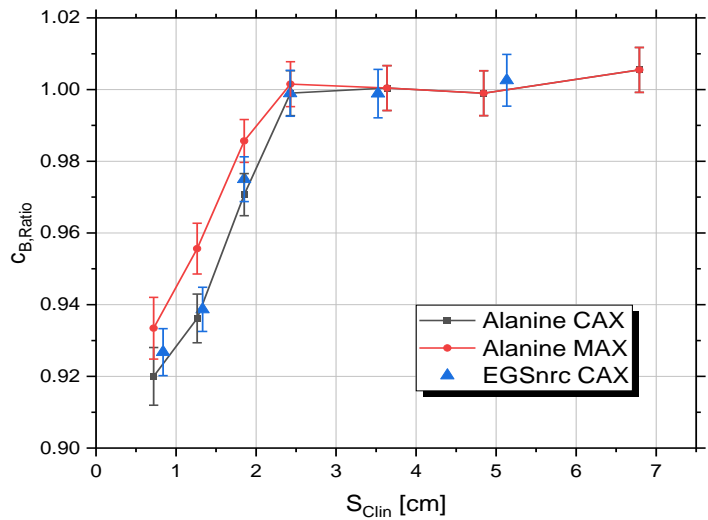

*Figure 4: $c_{\vec{B}_{\text{Ratio}}}$ including the standard uncertainty as a function of field size for the position at the central axis (CAX) and the profile's maximum (MAX), determined with alanine measurements with and without a 1.5 T magnetic field. The EGSnrc data taken from the previous work (Frick et al., 2025) is included for comparison.*

## 3.4 $k_{\vec{B},Q_{\text{Ratio}}}$ and $k_{\vec{B},Q_{\text{clin}},Q_{\text{msr}}}^{f_{\text{clin}},f_{\text{msr}}}$

In Figure 5, $k_{\vec{B},Q_{\text{Ratio}}}$ is shown and compared to available data given by (Blum et al., 2021; Cervantes et al., 2021; Frick et al., 2025; Margaroni et al., 2025). The $k_{\vec{B},Q_{\text{Ratio}}}$ data of this study is listed in the Supplementary Tables S.1 and S.2.

The combined standard uncertainty of $k_{\vec{B},Q_{\text{Ratio}}}$ was 0.013 (mean) and 0.015 (maximum) for the microDiamond and on average 0.011 (mean) and 0.022 (maximum) for the ionization chambers.

No significant differences of $k_{\vec{B},Q_{\text{Ratio}}}$ were found between the MAX and CAX positions for the microDiamond or between the MR-optimized version and their conventional counterpart.

For MAX, $k_{\vec{B},Q_{\text{Ratio}}}$ showed a linear decrease for all detectors towards field sizes <2.5 × 2.5 cm². For CAX, within the uncertainties $k_{\vec{B},Q_{\text{Ratio}}}$ remained constant over all field sizes for the PinPoint 3D-types and the Semiflex 3D-types in stem orientation (+1.5 T), while it decreased for small field sizes for the Semiflex 3D-types in the tip orientation (-1.5 T).

For the ionization chambers, $k_{\vec{B},Q_{\text{Ratio}}}$ determined at the MAX$_{\text{dose}}$ position did not follow the linear decreasing behaviour at the smallest field size. The mean difference between the MAX$_{\text{signal}}$ and MAX$_{\text{dose}}$ positions at the smallest field size was 0.021 ± 0.007.

The $k_{\vec{B},Q_{\text{Ratio}}}$ values were multiplied with $k_{Q_{\text{clin}},Q_{\text{msr}}}^{f_{\text{clin}},f_{\text{msr}}}$ values from the literature (Casar et al., 2020, 2019; TRS-483, 2017) to determine $k_{\vec{B},Q_{\text{clin}},Q_{\text{msr}}}^{f_{\text{clin}},f_{\text{msr}}}$, which is presented in Figure 6 including fits according to (TRS-483, 2017). The fit data is tabulated in the Supplementary Tables S.3 and S.4.

For the microDiamond, $k_{\vec{B},Q_{\text{clin}},Q_{\text{msr}}}^{f_{\text{clin}},f_{\text{msr}}}$ remained constant down to a field size of 2.5 × 2.5 cm² and decreased approximately linearly for smaller field sizes. For the Semiflex 3D types and the PinPoint 3D types, $k_{\vec{B},Q_{\text{clin}},Q_{\text{msr}}}^{f_{\text{clin}},f_{\text{msr}}}$ remained about constant down to approximately 2 × 2 cm².

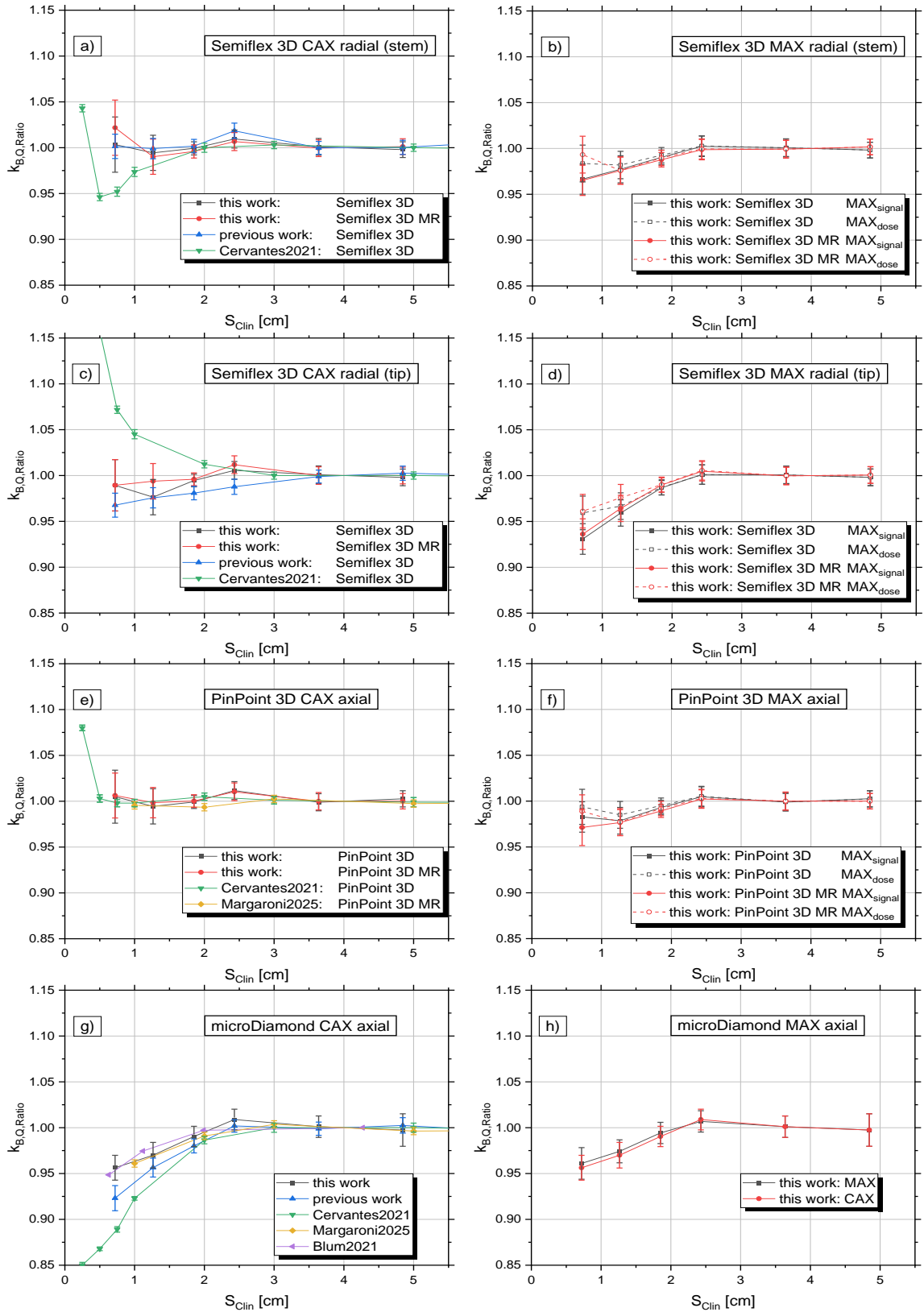

*Figure 5: $k_{\vec{B},Q_{Ratio}}$ including combined standard uncertainties for the CAX (left column) and MAX position (right column, see chapter 2.5) in a 1.5 T magnetic field for various detectors and orientations. For the ionization chambers at MAX, the reference point of the ionization chamber was positioned at the measured profiles maximum $MAX_{signal}$ and at the absorbed dose maximum $MAX_{dose}$ (determined by the microDiamond). For the microDiamond, these positions are the same. Like in this work, (Blum et al., 2021; Cervantes et al., 2021; Margaroni et al., 2025) – all of which employed Monte Carlo simulations - considers the volume effect of the detector.*

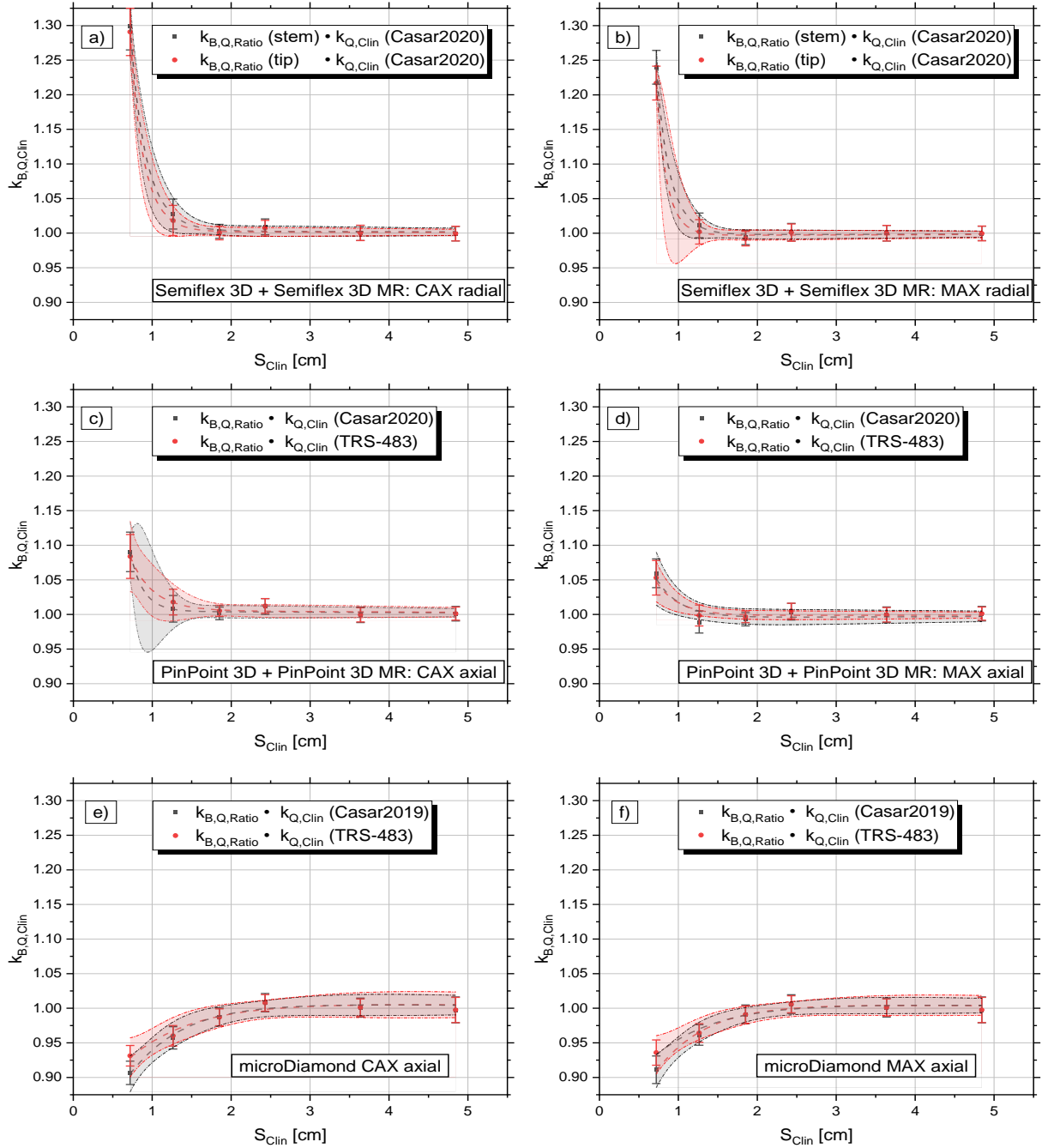

Figure 6: $k_{\vec{B},Q_{clin},Q_{msr}}^{f_{clin},f_{msr}}$ – including weighted least square fits with 95% confidence band determined according to (TRS-483, 2017) and calculated with the software Origin (OriginLab, USA) - determined with $k_{\vec{B},Q_{Ratio}}$ from this work and with $k_{Q_{clin},Q_{msr}}^{f_{clin},f_{msr}}$ values (including uncertainties) from the literature (Casar et al., 2020, 2019; TRS-483, 2017) for CAX (left column) and MAX position (right column, MAX$_{signal}$, see chapter 2.5) for different detectors and orientations. For the ionization chambers, the mean $k_{\vec{B},Q_{clin},Q_{msr}}^{f_{clin},f_{msr}}$ values of MR and non-MR chambers are presented. Due to fitting problems (non-convergent) for the fit of PinPoint 3D MAX, the "a" parameter was chosen manually.

# 4 Discussion

In this study, $k_{\vec{B},Q_{\text{Ratio}}}$ – representing the influence of the magnetic field on the output correction factor without a magnetic field ($k_{Q_{\text{clin}},Q_{\text{msr}}}^{f_{\text{clin}},f_{\text{msr}}}$) - and output correction factors, $k_{\vec{B},Q_{\text{clin}},Q_{\text{msr}}}^{f_{\text{clin}},f_{\text{msr}}}$, in a 1.5 T magnetic field were determined with a previously described method (Frick et al., 2025) fully experimentally for two MR-optimized chambers (Semiflex 3D MR PTW31024 and PinPoint 3D MR PTW31025), their conventional model (Semiflex 3D PTW31021 and PinPoint 3D PTW31022) and a solid-state detector (microDiamond PTW60019) for two positions: the central axis (CAX) and the maximum of the lateral dose profile (MAX).

## 4.1 General results

Since no difference in $k_{\text{pol}}$ was observed between measurements with and without a magnetic field for all field sizes, determining $k_{\text{pol}}$ separately for the method and detectors used is unnecessary, as it cancels out. This confirms the results of previous works (Frick et al., 2025).

The similar lateral profiles measured with the microDiamond and alanine indicate the suitability of the microDiamond for absorbed dose profile measurements in magnetic fields, as discussed in the literature (Episkopakis et al., 2024), where no need for an effective point of measurement (EPOM) shift was determined for the microDiamond. In the same publication, an average CAX-to-1.5 T-centre distance $d_{L,Centre}$ difference ($EPOM_{L,Centre}$ shift) between microDiamond or water voxel and four ionization chambers of 0.6 ± 0.1 mm was determined via Monte Carlo simulations, which was (within the given uncertainty) approx. the same for four ionization chambers and two detector orientations. This is the same value as the $EPOM_{L,MAX}$ shift in our study for all ionization chambers and approx. the same $EPOM_{L,Centre}$ shift for the PinPoint 3D types. In the same study and in (O'Brien et al., 2018), an experimental $d_{L,Centre}$ difference of about 1 mm between microDiamond and Semiflex 3D CAX shift was found, which is in the same range like in this work. Furthermore, (O'Brien et al., 2018) determined a $d_{L,Centre}$ difference between microDiamond and PinPoint 3D of only 0.2 mm, although a different detector orientation was used. While in (Episkopakis et al., 2024) inflection point shifts were investigated, in this study and in (O'Brien et al., 2018) 50% isodose shifts were analysed. Since the flattening filter free MR-linac profiles of small field sizes are very similar to profiles with a flattening filter (Frick et al., 2025), these shifts should be approx. the same.

With the values of these literature and from this work (different orientations), this leads to a mean $EPOM_{L,Centre}$ shift of 1.0 ± 0.2 mm (0.4 • radius) for the Semiflex 3D types and 0.4 ± 0.2 mm (0.25 • radius) for the PinPoint 3D types. In general, this shift seems to increase for larger ionization chambers (O'Brien et al., 2018).

For larger ionization chambers and non-axial positioning, the $EPOM_{L,Centre}$ shift and the $EPOM_{L,MAX}$ shift can differ, underlining the need to evaluate these shifts separately. Results of this work indicate that both $EPOM_{L,Centre}$ and $EPOM_{L,MAX}$ shifts are approx. constant over the field sizes.

In summary, the MAX position of the profiles measured with ionization chambers is not at the MAX position of the absorbed dose profiles because of the lateral EPOM shift. Since the maximum signal position is directly measurable, field output factors and $k_{\vec{B},Q_{\text{clin}},Q_{\text{msr}}}^{f_{\text{clin}},f_{\text{msr}}}$ for MAX does not rely on this lateral EPOM shift.

(O'Brien et al., 2018) determined no EPOM shift in depth (parallel to the beam axis) for the PinPoint 3D and a shift of -0.7 mm for the Semiflex 3D. This shift is important for depth dose curve measurements, but for relative profile measurements in small field sizes e.g. to determine the MAX position, such small shifts are negligible, as they would change the field width by e.g. only 0.09% with a shift of 1 mm (isocenter at 100 cm, Source-Detector-Distance of 110 cm).

## 4.2 $c_{\vec{B}_{\text{Ratio}}}$

Within the uncertainty, $c_{\vec{B}_{\text{Ratio}}}$ – representing the change in the absorbed dose to water due to the magnetic field as a function of field size – measured at CAX with alanine agrees with the simulated data from the previous work (Frick et al., 2025).

For field sizes below 2.5 × 2.5 cm² – which is approx. the field size for 6 MV when loss of lateral charged particle equilibrium (LCPE) occurs (TRS-483, 2017) -, $c_{\vec{B}_{\text{Ratio}}}$ decreases at both CAX and MAX, with a more pronounced decrease at CAX due to its lateral off-maximum position. This overall decline at both positions is primarily due to a larger difference in the percentage depth dose curve between no and a magnetic field at smaller field sizes, as demonstrated by (Frick et al., 2025) for CAX.

When neglecting scattered electrons that move non-parallel to the beam axis, the electron fluence at the measurement point in a magnetic field decreases in large fields because it is replaced by an electron fluence originating from deeper regions of the water phantom due to the curved electron paths. Photon intensity - and hence secondary-electron fluence - falls off exponentially with depth according to the Beer–Lambert law. In smaller fields, the electron fluence is further reduced since it originates from the penumbra of the profile (CAX vs. MAX).

In large fields with LCPE, the lateral scatter contribution with and without a magnetic field may stay roughly the same, which reduces the dose differences. As field size decreases and loss of LCPE occurs, the scatter contribution diminishes and dose differences grow.

At higher photon energies, secondary electrons have greater energy - and thus longer path lengths and larger gyroradii -, intensifying these effects and producing a more pronounced decrease in $c_{\vec{B}_{\text{Ratio}}}$.

## 4.3 $k_{\vec{B},Q_{\text{Ratio}}}$ and $k^{f_{\text{clin}},f_{\text{msr}}}_{\vec{B},Q_{\text{clin}},Q_{\text{msr}}}$

For the ionization chambers, $k_{\vec{B},Q_{\text{Ratio}}}$ is lower at MAX than at CAX. However, for the microDiamond, no significant differences were observed between CAX and MAX, indicating that data obtained for one position can be reliably used for the other or positions between.

$k_{\vec{B},Q_{\text{Ratio}}}$ remains constant down to 2.5 × 2.5 cm² for all detectors. For smaller field sizes at MAX, $k_{\vec{B},Q_{\text{Ratio}}}$ decreases linearly for all detectors. At CAX, for the Semiflex 3D type chambers (stem direction) and the PinPoint 3D types, $k_{\vec{B},Q_{\text{Ratio}}}$ appears to remain constant. This implies that $k^{f_{\text{clin}},f_{\text{msr}}}_{Q_{\text{clin}},Q_{\text{msr}}}$ can be used in the magnetic field without further correction. No significant difference in $k_{\vec{B},Q_{\text{Ratio}}}$ was observed between the novel MR-optimized chambers and their conventional model. At the smallest field size (0.7 × 0.7 cm²), $k_{\vec{B},Q_{\text{Ratio}}}$ of the ionization chambers positioned at MAX$_{\text{signal}}$ changed on average by 0.02 when shifted on average by 0.6 mm to the MAX$_{\text{dose}}$, underlining the high sensitivity of the positioning and therefore the high uncertainty of $k_{\vec{B},Q_{\text{Ratio}}}$ and $k^{f_{\text{clin}},f_{\text{msr}}}_{\vec{B},Q_{\text{clin}},Q_{\text{msr}}}$ at small field sizes.

The Semiflex 3D $k_{\vec{B},Q_{\text{Ratio}}}$ CAX values of this work significantly differs to values reported by (Cervantes et al., 2021), although discrepancy with these reported values were previously observed in other literature (Frick et al., 2025, 2024; Margaroni et al., 2025, 2023). Contrary, the constant PinPoint 3D $k_{\vec{B},Q_{\text{Ratio}}}$ values of this work agree within the uncertainties with values reported by (Cervantes et al., 2021; Margaroni et al., 2025). Furthermore, the general linear decreasing trend for $k_{\vec{B},Q_{\text{Ratio}}}$ values for the microDiamond matches with reported literature values (Cervantes et al., 2021; Frick et al., 2025; Margaroni et al., 2025).

$k^{f_{\text{clin}},f_{\text{msr}}}_{\vec{B},Q_{\text{clin}},Q_{\text{msr}}}$ remains constant down to a field size of 2.5 × 2.5 cm² for all detectors, confirming the key findings of the previous work (Frick et al., 2025). For ionization chambers at MAX, $k_{\vec{B},Q_{\text{Ratio}}}$ reduces the increasing $k^{f_{\text{clin}},f_{\text{msr}}}_{Q_{\text{clin}},Q_{\text{msr}}}$, leading to a need for a smaller correction. For example, for the PinPoint 3D types at the MAX position, $k^{f_{\text{clin}},f_{\text{msr}}}_{\vec{B},Q_{\text{clin}},Q_{\text{msr}}}$ remains approx. constant for longer down to a field size of 1.5 × 1.5 cm², indicating no need for correction down to this field size, and then increases only 5% at MAX instead of 9% at CAX.

The calculated TRS-483-based fits of $k^{f_{\text{clin}},f_{\text{msr}}}_{\vec{B},Q_{\text{clin}},Q_{\text{msr}}}$ exhibited problems (non-convergence or large confidence intervals) for datasets in which only a single data point at small field sizes deviated markedly from one (e.g. PinPoint 3D at CAX and MAX). This behaviour suggests that for field sizes below 1.5 × 1.5 cm², the determination of $k^{f_{\text{clin}},f_{\text{msr}}}_{\vec{B},Q_{\text{clin}},Q_{\text{msr}}}$ should be performed over finer field size intervals to ensure robust and reliable fits.

As described in (Cervantes et al., 2021), the output correction factor is an interplay between the overall perturbation of the detector, resulting from non-water components, and the volume-averaging effect. $k_{\vec{B},Q_{\text{Ratio}}}$ is constant for the PinPoint 3D types and Semiflex 3D types (stem direction) at CAX, while linearly decreasing at MAX for all detectors. This means that at CAX the ratio of the volume effect and overall perturbation effect in a magnetic field relative to a situation without a magnetic field stays constant or cancels each other out. For the PinPoint 3D and the microDiamond, the volume effect at CAX for the axial orientation used is approximately negligible for field sizes of ≥0.75 × 0.75 cm$^2$, with and without a magnetic field. Also, no change of the volume effect due to the magnetic field was reported for any field sizes (Cervantes et al., 2021).

Assuming the change of volume-averaging due to the magnetic field is constant over all field sizes, this means that the ratios of the perturbation of the PinPoint 3D types decrease at the MAX position while remaining approximately constant at CAX. Reasons for this may be that different detector parts and areas are irradiated at CAX and MAX. Furthermore for MAX, the product of fluence and electron path length in the sensitive volume is maximised, leading to an increased signal and therefore an overall reduction of the perturbation. For CAX, this is not the case since the EPOM shift is not considered, leading to a reduced signal, which may counter other increasing signal effects so that $k_{\vec{B},Q_{\text{Ratio}}}$ remains approx. constant.

For the microDiamond, this perturbation ratio decreases the same at both positions. As shown in (Blum et al., 2021), the high-density components of the microDiamond downstream of the sensitive volume are the main reason for the signal drop in a 1.5 T magnetic field of over 10% at big field sizes and about 5% in a small field. (Blum et al., 2021) speculated that for big fields the signal drop can be attributed to reduced backscattering from the enhanced density towards the sensitive volume. The decrease of this perturbation towards small field sizes was attributed to the decrease of lateral low energy secondary electrons reaching the detector, which are subject of stronger deflection. Another explanation is that in large fields, secondary electrons reach a detector at angles of roughly 0° ± 90° without a magnetic field and 45° ± 90° in a 1.5 T magnetic field (O'Brien and Sawakuchi, 2017; Tekin et al., 2022). Therefore, additionally to the reduced backscattering (Blum et al., 2021), the enhanced density downstream of the sensitive volume shields parts of the secondary electrons moving towards the sensitive volume from roughly 90° to 135°. In small field sizes when loss of LCPE occurs, this angle range is reduced, which may lead to a smaller shielding effect.

To gain a detailed understanding of the different perturbation components (extracameral components (stem, central electrode, cavity wall), medium and density), a study similar to (Cervantes et al., 2021) conducted at the maximum of the profile is required.

## 4.4 intra-type variation and comparison of methods

The standard uncertainty arising from intra-type variation for tabulated $k_{\vec{B},Q_{\text{Ratio}}}$ and $k_{\vec{B},Q_{\text{clin}},Q_{\text{msr}}}^{f_{\text{clin}},f_{\text{msr}}}$ is non-neglectable (up to 0.009) and therefore must be considered. Since the maximum uncertainty of the detector output ratios was approximately the same with and without a magnetic field, the intra-type variation uncertainty of $k_{Q_{\text{clin}},Q_{\text{msr}}}^{f_{\text{clin}},f_{\text{msr}}}$ can be used as a good estimate for $k_{\vec{B},Q_{\text{clin}},Q_{\text{msr}}}^{f_{\text{clin}},f_{\text{msr}}}$.

The maximum difference of $k_{\vec{B},Q_{\text{Ratio}}}$ compared to values from previous work (Frick et al., 2025) (conducted over one year apart) was 0.03 found at the smallest field size. This difference can be attributed to intra-type variation, different $c_{\vec{B}_{\text{Ratio}}}$, and reproducibility (including the long-term stability of the measurement devices), with the latter being the primary contributor.

This maximum difference was higher with 0.07 when directly determining $k_{\vec{B},Q_{\text{clin}},Q_{\text{msr}}}^{f_{\text{clin}},f_{\text{msr}}}$ (Method A from previous work), confirming that this method ($k_{\vec{B},Q_{\text{Ratio}}}$) may be more advantageous to cope with the high uncertainty of the output correction factors for small field sizes.

### 4.5 Advantages of the CAX and MAX positioning

Opposed to an ideal situation without a magnetic field, in a magnetic field the maximum and centre of the measured profiles as well as the central axis of the photon beam are different for small field sizes and therefore a measurement position must be chosen. All positions have its own advantages.

On the one hand, CAX data are measured quickly as the detector is not moved, eliminating uncertainties or user errors in determining the field maximum or the centre, which is a drawback of the maximum and centre measuring positions. On the other hand, for small field sizes, the field's centre or maximum can shift slightly even without a magnetic field, as was the case in the experimental setup of this work.

Furthermore, for general tabulated factors, the measurement position at the penumbra of the profile at smaller field sizes may suggest that comparable profiles are needed for comparable results between laboratories. Otherwise, larger differences must be accepted, leading to higher uncertainties in the tabulated factors. Therefore, measuring at the maximum could lead to more comparable factors between the results in the literature.

At an MR-linac, the MAX and the centre position are directly measurable and thus allows for a precise detector placement. In contrast, the CAX position relies on position placement with e.g. the Truefix system (PTW) or MV-imaging, since in clinical routine the magnetic field cannot be turned off. In a recent multi-centre study, the shift between CAX and the centre of the dose profile measured with different detectors has been investigated for reference fields at five different 1.5 T MR-linacs, including different serial numbers of the used detector type per MR-linac (Episkopakis et al., 2024). For the microDiamond, a mean CAX deviation of 1.25 mm±0.07 mm was found, which is also in the range of our reported value measured at an MR-linac from the previous work, which was approximately constant across all field sizes (1.33 mm ± 0.10 mm) (Frick et al., 2025).

This underlines the generalizability of the CAX deviation and therefore the stability of the CAX positioning, emphasising the usability of general tabulated $k_{\vec{B},Q_{\text{clin}},Q_{\text{msr}}}^{f_{\text{clin}},f_{\text{msr}}}$ for the CAX positioning for a 1.5 T MR-linac. For very small field sizes, the more precise MAX positioning and for ionization chambers a smaller $k_{\vec{B},Q_{\text{clin}},Q_{\text{msr}}}^{f_{\text{clin}},f_{\text{msr}}}$ may be advantageous.

## 5  Conclusion

This work determined $k_{\vec{B},Q_{\text{Ratio}}}$ - the influence of the magnetic field on the output correction factor without a magnetic field ($k_{Q_{\text{clin}},Q_{\text{msr}}}^{f_{\text{clin}},f_{\text{msr}}}$) – and the output correction factor $k_{\vec{B},Q_{\text{clin}},Q_{\text{msr}}}^{f_{\text{clin}},f_{\text{msr}}}$ in a 1.5 T magnetic field fully experimentally for the central axis (CAX) and the maximum of the measured lateral profile (MAX) for two novel MR-optimized ionization chambers (sensitive volume 0.07/0.016 cm³), their conventional counterpart, and a solid-state detector for square field sizes ($S_{\text{clin}}$) down to 0.7 × 0.7 cm². Furthermore, the uncertainty of the intra-type variation for tabulated $k_{\vec{B},Q_{\text{Ratio}}}$ and $k_{\vec{B},Q_{\text{clin}},Q_{\text{msr}}}^{f_{\text{clin}},f_{\text{msr}}}$ was assessed. The absorbed dose to water was measured with alanine/EPR.

The reference point of the ionization chambers' signal maximum position was on average for all field sizes at a lateral distance of 0.6 ± 0.1 mm compared to the dose maximum position. For all detectors, $k_{\vec{B},Q_{\text{Ratio}}}$ at both CAX and MAX remained constant down to a field size of 2.5 × 2.5 cm². At MAX for field sizes below 2.5 × 2.5 cm², $k_{\vec{B},Q_{\text{Ratio}}}$ decreased linearly for all detectors. No significant difference of $k_{\vec{B},Q_{\text{Ratio}}}$ was found between CAX and MAX for the solid-state detector, indicating that $k_{\vec{B},Q_{\text{clin}},Q_{\text{msr}}}^{f_{\text{clin}},f_{\text{msr}}}$ data obtained for one position can be reliably used for the other. At CAX, for the 0.07 cm³ ionization chambers in the stem orientation (electrons moving along the beam axis are deflected towards the stem due to the Lorentz Force) and for the 0.016 cm³ ionization chambers, $k_{\vec{B},Q_{\text{Ratio}}}$ remained approximately constant, indicating the usability of $k_{Q_{\text{clin}},Q_{\text{msr}}}^{f_{\text{clin}},f_{\text{msr}}}$ without further correction.

Furthermore, no significant difference of $k_{\vec{B},Q_{\text{Ratio}}}$ between the MR-optimized chambers and their counterparts were observed, indicating that the same values of $k_{\vec{B},Q_{\text{clin}},Q_{\text{msr}}}^{f_{\text{clin}},f_{\text{msr}}}$ can be used for both chamber types.

$k_{\vec{B},Q_{\text{clin}},Q_{\text{msr}}}^{f_{\text{clin}},f_{\text{msr}}}$ remained constant down to a field size of approx. 2.5 × 2.5 cm² for all detectors, orientations and positions. For smaller field sizes, $k_{\vec{B},Q_{\text{clin}},Q_{\text{msr}}}^{f_{\text{clin}},f_{\text{msr}}}$ increased up to 1.28 for the 0.07 cm³-chamber types and 1.09 for the 0.016 cm³-chamber types. For ionization chambers, $k_{\vec{B},Q_{\text{clin}},Q_{\text{msr}}}^{f_{\text{clin}},f_{\text{msr}}}$ was smaller at MAX than at CAX. For the solid-state detector, $k_{\vec{B},Q_{\text{clin}},Q_{\text{msr}}}^{f_{\text{clin}},f_{\text{msr}}}$ decreased up to 0.92 at both MAX and CAX.

The standard uncertainty for tabulated $k_{\vec{B},Q_{\text{Ratio}}}$ and $k_{\vec{B},Q_{\text{clin}},Q_{\text{msr}}}^{f_{\text{clin}},f_{\text{msr}}}$ arising from intra-type variation is non-neglectable (up to 0.009) for both factors and therefore must be considered.

The result of this work underlines the necessity to define a position for $k_{\vec{B},Q_{\text{clin}},Q_{\text{msr}}}^{f_{\text{clin}},f_{\text{msr}}}$ for ionization chambers. For general tabulated $k_{\vec{B},Q_{\text{clin}},Q_{\text{msr}}}^{f_{\text{clin}},f_{\text{msr}}}$ and thus for field output factor measurements, both CAX and MAX positions are useable. For very small field sizes, the more precise MAX positioning and for ionization chambers a smaller $k_{\vec{B},Q_{\text{clin}},Q_{\text{msr}}}^{f_{\text{clin}},f_{\text{msr}}}$ may be advantageous.

## 7 Acknowledgement

This study received funding by the EMPIR program (Grant 19NRM01 MRgRT-DOS). We thank PTW Freiburg for providing microDiamond detectors for this work.

# 8 Supplementary Material

## 8.1 Profile Comparison: microDiamond vs. alanine

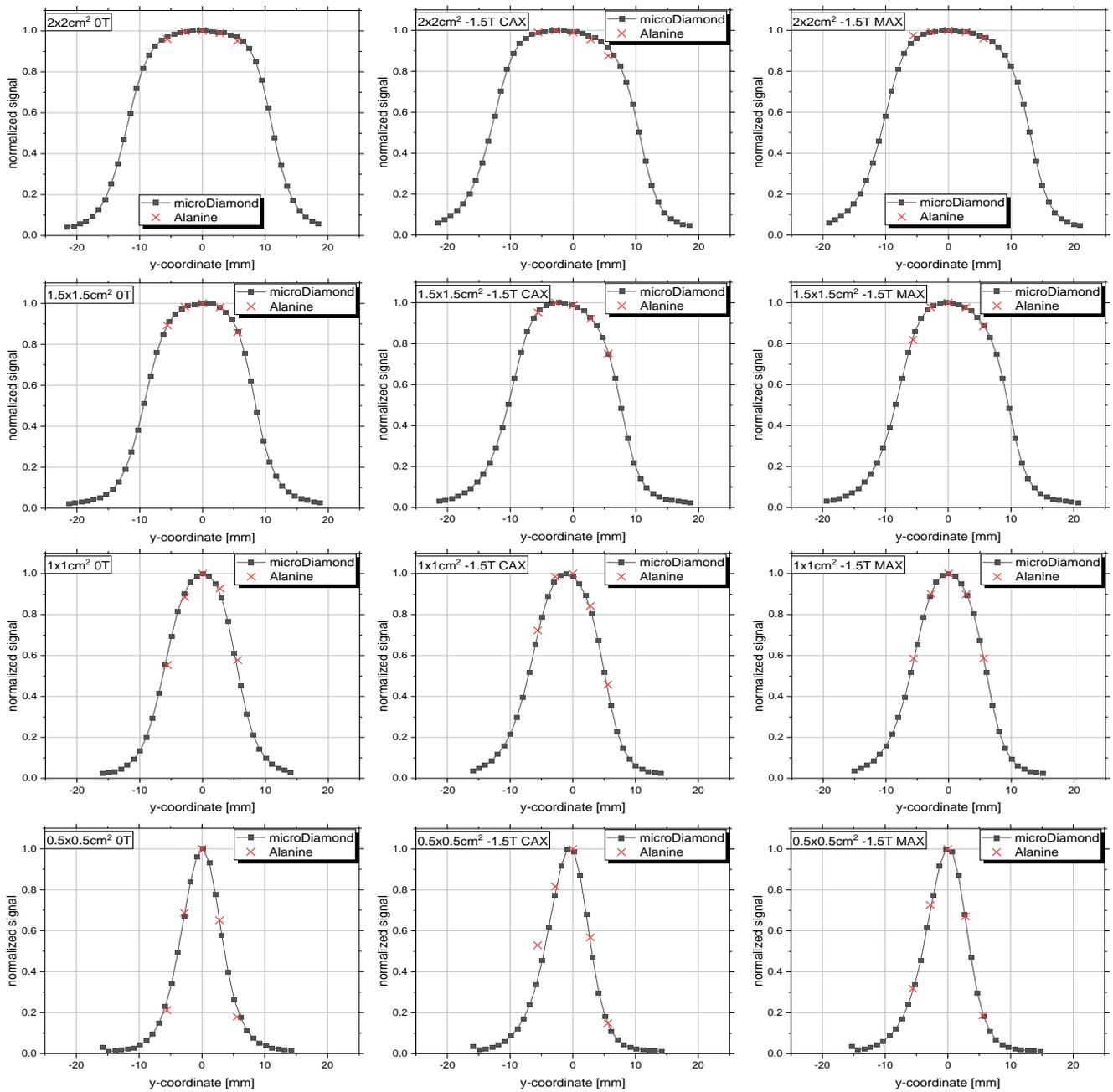

*Figure S.1: Comparison of profiles measured with a microDiamond detector and alanine pellets without a magnetic field (left column) and with a 1.5 T magnetic field at CAX (middle column) and at MAX (right column) for different field sizes.*

## 8.2 Tabulated $k_{\vec{B},Q_{\text{Ratio}}}$ values

In Table S.1 (CAX) and S.2 (MAX), $k_{\vec{B},Q_{\text{Ratio}}}$ data presented in Figure 5 including their combined standard uncertainty (see chapter 2.3 and chapter 2.3) are listed. Like in the figure, the data is normalized to $S_{\text{clin}}$ = 4 cm.

*Table S.1: $k_{\vec{B},Q_{Ratio}}$ at 1.5 T for CAX presented in Figure 5, including combined standard uncertainties.*

| $k_{\vec{B},Q_{\text{Ratio}}}$ CAX | micro-Diamond (axial) | Semiflex 3D (radial) | | Semiflex 3D MR (radial) | | PinPoint 3D (axial) | PinPoint 3D MR (axial) |
|---|---|---|---|---|---|---|---|
| $S_{\text{clin}}$ [cm] | | stem | tip | stem | tip | | |
| 4.84 | 0.9974(178) | 0.9979(86) | 0.9979(91) | 1.0012(86) | 1.0001(91) | 1.0026(88) | 1.0000(84) |
| 3.64 | 1.0011(117) | 1.0009(92) | 1.0009(96) | 0.9995(92) | 1.0000(96) | 0.9989(94) | 1.0000(95) |
| 2.43 | 1.0089(113) | 1.0095(98) | 1.0056(97) | 1.0067(98) | 1.0118(97) | 1.0116(98) | 1.0103(97) |
| 1.85 | 0.9905(111) | 0.9993(72) | 0.9946(69) | 0.9959(72) | 0.9960(69) | 0.9992(71) | 1.0003(69) |
| 1.26 | 0.9700(140) | 0.9944(192) | 0.9764(193) | 0.9901(192) | 0.9938(193) | 0.9944(192) | 0.9985(166) |
| 0.72 | 0.9563(136) | 1.0034(301) | 0.9893(280) | 1.0218(301) | 0.9894(280) | 1.0050(289) | 1.0063(245) |

*Table S.2: $k_{\vec{B},Q_{Ratio}}$ at 1.5 T for MAX presented in Figure 5, including combined standard uncertainties.*

| $k_{\vec{B},Q_{\text{Ratio}}}$ MAX | micro-Diamond (axial) | Semiflex 3D (radial) | | Semiflex 3D MR (radial) | | PinPoint 3D (axial) | PinPoint 3D MR (axial) |
|---|---|---|---|---|---|---|---|
| $S_{\text{clin}}$ [cm] | | stem | tip | stem | tip | | |
| 4.84 | 0.9974(178) | 0.9982(86) | 0.9981(91) | 1.0017(86) | 1.0006(91) | 1.0026(88) | 1.0000(84) |
| 3.64 | 1.0011(117) | 1.0008(96) | 1.0008(95) | 0.9993(96) | 0.9997(95) | 0.9989(95) | 1.0000(97) |
| 2.43 | 1.0069(116) | 1.0023(111) | 1.0012(107) | 0.9987(111) | 1.0048(107) | 1.0047(109) | 1.0023(101) |
| 1.85 | 0.9943(116) | 0.9903(79) | 0.9868(78) | 0.9877(79) | 0.9898(78) | 0.9930(79) | 0.9893(70) |
| 1.26 | 0.9742(125) | 0.9773(148) | 0.9593(143) | 0.9759(148) | 0.9642(143) | 0.9786(145) | 0.9765(143) |
| 0.72 | 0.9610(172) | 0.9666(165) | 0.9309(167) | 0.9653(165) | 0.9362(167) | 0.9827(166) | 0.9713(198) |

## 8.3 Tabulated $k^{f_{\text{clin}},f_{\text{msr}}}_{\vec{B},Q_{\text{clin}},Q_{\text{msr}}}$ fit values

In Table S.3 (CAX) and S.4 (MAX), $k^{f_{\text{clin}},f_{\text{msr}}}_{\vec{B},Q_{\text{clin}},Q_{\text{msr}}}$ fit value data from Figure 6 including their 68% confidence interval as standard uncertainty are presented, normalized to $S_{\text{clin}}$ = 4 cm. $k^{f_{\text{clin}},f_{\text{msr}}}_{\vec{B},Q_{\text{clin}},Q_{\text{msr}}}$ was calculated using $k_{\vec{B},Q_{\text{Ratio}}}$ determined in this work (Figure 5) and $k^{f_{\text{clin}},f_{\text{msr}}}_{Q_{\text{clin}},Q_{\text{msr}}}$ values from the literature (Casar et al., 2020, 2019; TRS-483, 2017).

Table S.3: $k^{f_{\text{clin}},f_{\text{msr}}}_{\vec{B},Q_{\text{clin}},Q_{\text{msr}}}$ fit values for 1.5 T and CAX presented in Figure 6, including their 68% confidence intervals and normalized to $S_{\text{clin}}$ = 4 cm.

| $k^{f_{\text{clin}},f_{\text{msr}}}_{\vec{B},Q_{\text{clin}},Q_{\text{msr}}}$ CAX | microDiamond (axial) | | Semiflex 3D + Semiflex 3D MR (radial) | | PinPoint 3D + PinPoint 3D MR (axial) | |
|---|---|---|---|---|---|---|
| $S_{\text{clin}}$ [cm] | $k_{\vec{B},Q_{\text{Ratio}}} \cdot k^{f_{\text{clin}},f_{\text{msr}}}_{Q_{\text{clin}},Q_{\text{msr}}}$ (Casar2019) | $k_{\vec{B},Q_{\text{Ratio}}} \cdot k^{f_{\text{clin}},f_{\text{msr}}}_{Q_{\text{clin}},Q_{\text{msr}}}$ (TRS-483) | $k_{\vec{B},Q_{\text{Ratio}}}$ (stem) $\cdot$ $k^{f_{\text{clin}},f_{\text{msr}}}_{Q_{\text{clin}},Q_{\text{msr}}}$ (Casar2020) | $k_{\vec{B},Q_{\text{Ratio}}}$ (tip) $\cdot$ $k^{f_{\text{clin}},f_{\text{msr}}}_{Q_{\text{clin}},Q_{\text{msr}}}$ (Casar2020) | $k_{\vec{B},Q_{\text{Ratio}}} \cdot k^{f_{\text{clin}},f_{\text{msr}}}_{Q_{\text{clin}},Q_{\text{msr}}}$ (Casar2020) | $k_{\vec{B},Q_{\text{Ratio}}} \cdot k^{f_{\text{clin}},f_{\text{msr}}}_{Q_{\text{clin}},Q_{\text{msr}}}$ (TRS-483) |
| 4.00 | 1.0000(78) | 1.0000(96) | 1.0000(32) | 1.0000(26) | 1.0000(36) | 1.0000(41) |
| 3.00 | 0.9985(70) | 0.9977(78) | 0.9997(37) | 0.9997(31) | 1.0005(41) | 1.0006(47) |
| 2.50 | 0.9953(61) | 0.9941(69) | 0.9996(38) | 0.9995(32) | 1.0007(44) | 1.0011(47) |
| 2.00 | 0.9877(64) | 0.9868(77) | 0.9998(37) | 0.9993(32) | 1.0010(45) | 1.0022(43) |
| 1.50 | 0.9708(88) | 0.9728(96) | 1.0030(65) | 1.0000(51) | 1.0021(59) | 1.0069(83) |
| 1.00 | 0.9349(93) | 0.9468(98) | 1.0484(234) | 1.0268(244) | 1.0174(367) | 1.0311(192) |
| 0.70 | 0.8978(133) | 0.9222(145) | 1.2753(217) | 1.2602(182) | 1.0990(218) | 1.0858(261) |

Table S.4: $k^{f_{\text{clin}},f_{\text{msr}}}_{\vec{B},Q_{\text{clin}},Q_{\text{msr}}}$ fit values for 1.5 T and MAX presented in Figure 6, including their 68% confidence intervals and normalized to $S_{\text{clin}}$ = 4 cm.

| $k^{f_{\text{clin}},f_{\text{msr}}}_{\vec{B},Q_{\text{clin}},Q_{\text{msr}}}$ MAX | microDiamond (axial) | | Semiflex 3D + Semiflex 3D MR (radial) | | PinPoint 3D + PinPoint 3D MR (axial) | |
|---|---|---|---|---|---|---|
| $S_{\text{clin}}$ [cm] | $k_{\vec{B},Q_{\text{Ratio}}} \cdot k^{f_{\text{clin}},f_{\text{msr}}}_{Q_{\text{clin}},Q_{\text{msr}}}$ (Casar2019) | $k_{\vec{B},Q_{\text{Ratio}}} \cdot k^{f_{\text{clin}},f_{\text{msr}}}_{Q_{\text{clin}},Q_{\text{msr}}}$ (TRS-483) | $k_{\vec{B},Q_{\text{Ratio}}}$ (stem) $\cdot$ $k^{f_{\text{clin}},f_{\text{msr}}}_{Q_{\text{clin}},Q_{\text{msr}}}$ (Casar2020) | $k_{\vec{B},Q_{\text{Ratio}}}$ (tip) $\cdot$ $k^{f_{\text{clin}},f_{\text{msr}}}_{Q_{\text{clin}},Q_{\text{msr}}}$ (Casar2020) | $k_{\vec{B},Q_{\text{Ratio}}} \cdot k^{f_{\text{clin}},f_{\text{msr}}}_{Q_{\text{clin}},Q_{\text{msr}}}$ (Casar2020) | $k_{\vec{B},Q_{\text{Ratio}}} \cdot k^{f_{\text{clin}},f_{\text{msr}}}_{Q_{\text{clin}},Q_{\text{msr}}}$ (TRS-483) |
| 4.00 | 1.0000(60) | 1.0000(76) | 1.0000(26) | 1.0000(28) | 1.0000(45) | 1.0000(26) |
| 3.00 | 0.9989(56) | 0.9982(64) | 0.9997(31) | 0.9997(33) | 0.9995(53) | 0.9998(31) |
| 2.50 | 0.9963(49) | 0.9951(56) | 0.9996(32) | 0.9995(36) | 0.9994(56) | 0.9999(32) |
| 2.00 | 0.9898(49) | 0.9887(61) | 0.9998(33) | 0.9993(37) | 0.9998(57) | 1.0003(33) |
| 1.50 | 0.9744(67) | 0.9759(76) | 1.0030(42) | 1.0000(40) | 1.0032(57) | 1.0030(34) |
| 1.00 | 0.9399(76) | 0.9510(85) | 1.0484(202) | 1.0268(343) | 1.0210(98) | 1.0182(70) |
| 0.70 | 0.9025(122) | 0.9267(142) | 1.2753(125) | 1.2602(141) | 1.0592(195) | 1.0519(147) |